\documentstyle[eqsecnum,aps]{revtex}

\begin{document}
\draft \preprint{HEP/123-qed}
\title{Phonon Thermodynamics versus Electron-Phonon Models }
\author{Marco Zoli}
\address{Istituto Nazionale Fisica della Materia - Dipartimento di Fisica - Universit\'a di Camerino, \\
62032 Italy. e-mail: marco.zoli@unicam.it }

\date{\today}
\maketitle
\begin{abstract}
Applying the path integral formalism we study the equilibrium
thermodynamics of the phonon field both in the Holstein and in the
Su-Schrieffer-Heeger models. The anharmonic cumulant series,
dependent on the peculiar source currents of the {\it e-ph}
models, have been computed versus temperature in the case of a low
energy oscillator. The cutoff in the series expansion has been
determined, in the low $T$ limit, using the constraint of the
third law of thermodynamics. In the Holstein model, the free
energy derivatives do not show any contribution ascribable to {\it
e-ph} anharmonic effect. We find signatures of large {\it e-ph}
anharmonicities in the Su-Schrieffer-Heeger model mainly visible
in the temperature dependent peak displayed by the phonon heat
capacity.
\end{abstract} 

\narrowtext

\section*{I. Introduction}

The equilibrium thermodynamical properties of real systems provide
signatures of phonon anharmonicities induced both by
electron-phonon and phonon-phonon interactions. While in general
the exact thermodynamics of anharmonic {\it e-ph} model
Hamiltonians can be derived only in one dimension and in the
classical limit, variational methods  based on the Bogoliubov's
inequality \cite{falk} permit to obtain {\it i)} upper bounds for
the partition functions \cite{raedt} and {\it ii)} a correct
estimate of the entropy within the harmonic theory once the
harmonic phonons are self-consistently replaced by the temperature
dependent phonon frequencies \cite{huiallen}. Such a replacement
however does not hold for other thermodynamic functions and
constant volume anharmonic effects have to be explicitly computed
by diagrammatic perturbative methods \cite{barron,io1}. As at high
temperatures phonon-phonon anharmonicities are large, in the
intermediate to low temperature range the {\it e-ph} coupling
strength may induce the dominant anharmonic effects in the phonon
subsystem, the latter being dependent on the peculiarities of the
{\it e-ph} model Hamiltonian.

The Holstein Hamiltonian \cite{holst} and the Su-Schriffer-Heeger
Hamiltonian \cite{ssh} have been widely investigated in the last
decades also in view of the growing interest for polaronic
mechanisms in high $T_c$ superconductors \cite{alemott} and
polymers \cite{lu}. While the Holstein model assumes a local
coupling of tight binding electrons to optical phonons, in the SSH
model the electron-phonon interaction modifies the electron
hopping matrix elements thus leading to a non-local (in momentum
space) coupling with vertex function dependent on both the
electron and the phonon wave vector \cite{sto,per}. As the ground
state of the SSH Hamiltonian is twofold degenerate localized
solitonic solutions appear in the system together with polarons
\cite{campbell,stafstrom} whose formation depends however on the
strength of the electron coupling to the induced lattice
deformation \cite{ono} and on the value of the adiabatic parameter
\cite{io2}.

Analysis of the Holstein and SSH Hamiltonians generally assume
harmonic phonon models although lattice anharmonicities may induce
relevant microscopic effects such as the reduction of the Holstein
polaron effective mass \cite{voulga} along with a consistent
broadening of the polaron size \cite{chris}. In large polaron
transport models based on the Holstein Hamiltonian, the effects of
the phonon spectrum renormalization on the polaron scattering have
been studied  both for the optical \cite{ht} and the acoustical
\cite{sh} branches. Phonon anharmonicities have been also proposed
as a fundamental feature favouring high $T_c$ superconductivity in
the context of (bi)polaronic theories \cite{emin,mahan,zhong}.

When the electron energy scale, set by the value of the tight
binding overlap integral, is larger of (or comparable to) the
characteristic phonon energy the lattice deformation does not
follow instantaneously the electron motion \cite{atin} and the
{\it e-ph} interaction becomes time dependent. The path integral
formalism \cite{feynman} is suitable to account for time dependent
interactions as a retarded potential naturally emerges in the
exact integral action \cite{kleinert}. Moreover, being valid for
any {\it e-ph} coupling value, the path integral method allows one
to derive the partition function and the related temperature
derivatives without those limitations which affect the
perturbative studies.

Our previous investigations have focussed on the equilibrium
thermodynamics of the SSH model by assuming a bath of {\it
harmonic} phonons for the electron particle path \cite{io3} while
anharmonic corrections to the phonon free energy had been only
partially considered \cite{io4} in the case of a time averaged
electron particle path. However, being the {\it e-ph} SSH
interaction intrinsecally non local on the time scale, significant
phonon anharmonicities may be found through a full computation of
the cumulant expansion depending on the perturbing source current
of the Hamiltonian model. With the present paper we propose a
comparative analysis of the equilibrium thermodynamics for the
phonon system in the Holstein and SSH models pointing out the
appearance of macroscopic anharmonic features driven by the
strength of the {\it e-ph} coupling. In Section II we outline the
Hamiltonian models and the cumulant expansion for the phonon
partition function. The results are presented in Section III,
while Section IV contains some final remarks.

\section*{II. The Hamiltonian Models}

In one dimension the {\it e-ph} Holstein Hamiltonian ($H^H$) and
the {\it e-ph} SSH Hamiltonian ($H^{SSH}$) read respectively:

\begin{eqnarray}
& &H^{H}=\, -{J \over 2} \sum_{r} \bigl( f^{\dag}_r f_{r+1} +
f^{\dag}_{r+1} f_{r} \bigr) + g \sum_{r} f^{\dag}_r f_{r} \bigl(
b^{\dag}_r + b_r \bigr)  \, \nonumber \\ & &H^{SSH}=\,
\sum_{r}J_{r,r+1} \bigl(f^{\dag}_r f_{r+1} + f^{\dag}_{r+1} f_{r}
\bigr) \, \nonumber \\ & &J_{r,r+1}=\, - {1 \over 2}\bigl[ J +
\alpha (u_r - u_{r+1})\bigr] \, \nonumber \\
\end{eqnarray}

where, $g$ is the Holstein {\it e-ph} coupling in {\it energy}
units, $\alpha$ is the SSH {\it e-ph} coupling in units {\it
energy $\times$ length$^{-1}$}, $J$ is the nearest neighbors
hopping integral for an undistorted chain, $f^{\dag}_r$ and
$f_{r}$ create and destroy electrons on the $r-$ lattice site,
$u_r$ is the displacement of the $r-th$ atomic group along the
molecular axis, $b^{\dag}_r$ and $b_{r}$ are the real space
creation and annihilation phonon operators. The non-interacting
phonon Hamiltonian $H_{ph}$ is assumed to be given, in both
models, by a set of independent oscillators. Taking $\omega$ as
the characteristic phonon energy, the displacement field can be
written as $u_r=\,(2M \omega)^{-1/2} \bigl(b^{\dag}_r + b_{r}
\bigr)$ and $H^H$ transforms into

\begin{eqnarray}
H^{H}=\,& & -{J \over 2} \sum_{r} \bigl( f^{\dag}_r f_{r+1} +
f^{\dag}_{r+1} f_{r} \bigr) + {\bar g} \sum_{r}u_r f^{\dag}_r
f_{r}  \, \nonumber \\ {\bar g}=\, & & g \sqrt{2M \omega}
\end{eqnarray}

with ${\bar g}$ dimensionally equivalent to the coupling $\alpha$
of the SSH model. Choosing the atomic mass $M$ of order $10^4$
times the electron mass, we get ${\bar g} \simeq \, 1.1456 \times
g \sqrt{2\omega}{}\, \bigl[ meV \AA^{-1} \bigr]$ where $g$ is
given in units of $\omega$ and the latter is given in $meV$. This
transformation allows us to compare numerically the {\it e-ph}
effective coupling in the two Hamiltonian models.

Let's apply to the Hamiltonians in eq.(1) space-time mapping
techniques previously used i.e. in the path integral formulation
of the theory of dilute magnetic alloys \cite{hamann} and in the
path integral study of A15 compounds with strong electron-phonon
coupling \cite{yuand}.

Thus, we introduce $x(\tau)$ and $y(\tau')$ as the electron
coordinates at the $r$ and $r+1$ lattice sites respectively, while
the spatial {\it e-ph} correlations contained in (1) are mapped
onto the time axis by changing: $u_r \to u(\tau)$ and $u_{r+1} \to
u(\tau')$. Then, $H^H$ in (2) and $H^{SSH}$ in (1) transform to

\begin{eqnarray}
H^{H}(\tau,\tau'& &)=\, -{J \over 2} \bigl( f^{\dag}(x(\tau))
f(y(\tau')) + f^{\dag}(y(\tau')) f(x(\tau)) \bigr) \, \nonumber
\\ & &{}+  {\bar g} u(\tau) f^{\dag}(x(\tau)) f(x(\tau)) \, \nonumber
\\ H^{SSH}(\tau,& &\tau')=\, J_{\tau, \tau'}
\Bigl(f^{\dag}(x(\tau)) f(y(\tau')) + f^{\dag}(y(\tau'))
f(x(\tau)) \Bigr)\, \nonumber \\ & &J_{\tau, \tau'}=\, - {1 \over
2}\bigl[J + \alpha(u(\tau) - u(\tau'))\bigr]
\end{eqnarray}

$\tau$ and $\tau'$ are continuous variables $\bigl( \in [0,
\beta]\bigr)$ in the Matsubara Green's functions formalism with
$\beta$ being the inverse temperature hence, the electron hops are
not constrained to first neighbors sites along the chain.
Accordingly, the Hamiltonians in (3) are more general than the
real space Hamiltonians in (1). Note that, while the Holstein
model assumes a local {\it e-ph} interaction along the $\tau-$
axis, in the SSH picture the electron hopping is accompanied by a
time dependent displacement of the atomic coordinate.

The ground state of the SSH Hamiltonian is twofold degenerate and,
in real space, a soliton connects the two phases with different
senses of dimerization. As a localized electronic state is
associated to the soliton the SSH model describes in principle
both electron hopping between solitons and thermally activated
hopping to band states. By mapping onto the $\tau-$ axis we
introduce time dependent electron hops while maintaining the
fundamental features of the SSH Hamiltonian. Eqs.(3) display the
semiclassical nature of our method in which quantum mechanical
degrees of freedom interact with the classical variables
$u(\tau)$. After setting $\tau'=\,0$, $u(0)\equiv y(0) \equiv 0$,
we take the thermal averages for the electron operators over the
ground states of the respective Hamiltonians thus obtaining the
average energies per lattice site due to electron hopping plus
{\it e-ph} coupling:

\begin{eqnarray}
{{<H^H(\tau)>} \over N}=& &\, - {J \over
2}G^{\pm}\bigl[x(\tau),\tau \bigr] + {\bar g}u(\tau) \,\nonumber
\\ {{<H^{SSH}(\tau)>} \over N}=& &\, - \Bigl({J \over 2} + \alpha
u(\tau) \Bigr) G^{\pm}\bigl[x(\tau),\tau \bigr] \,\nonumber \\
G^{\pm}\bigl[x(\tau),\tau \bigr]=& &\,G[-x(\tau), -\tau ] +
G[x(\tau), \tau ] \,\nonumber \\
=& &\,{2a \over
{\pi}}\int_{0}^{\pi/a}{dk} \cos[k x(\tau)]
\cosh\bigl(\varepsilon_k \tau \bigr) n_F(\varepsilon_k)\,\nonumber
\\
\end{eqnarray}

where $N$ is the number of lattice sites, $G[x(\tau), \tau ]$ is
the electron propagator, $a$ is the lattice constant,
$\varepsilon_k=\, - {J \over 2}\cos(ka)$ and $n_F(\varepsilon_k)$
is the Fermi function.

In general, the phonon partition function perturbed by a source
current $j(\tau)$ can be expanded in anharmonic series as:

\begin{eqnarray}
& &Z_{ph}[j(\tau)]\simeq \, Z_{h} \biggl(1 + \sum_{l=1}^k (-1)^l
<C^l>_{j(\tau)} \biggr)  \, \nonumber \\ & &Z_{h}=\,
\prod_{i=1}^{N} {1 \over {2\sinh(\omega_i\beta/2)}}
\end{eqnarray}

where the cumulant terms  $<C^l>_{j(\tau)}$ are expectation values
of powers of correlation functions of the perturbing current. The
averages are meant over the ensemble of the $N$ harmonic
oscillators (having energies $\omega_i$) whose partition function
is $Z_{h}$. As the oscillators are decoupled we study the
anharmonicity induced on a single oscillator with energy $\omega$
by the source currents peculiar of the Hamiltonians in (3). Being,
$j^H(\tau)=\,{\bar g}u(\tau)$ and $j^{SSH}(\tau)=\,- \alpha
u(\tau)G^{\pm}\bigl[x(\tau),\tau \bigr]$ we get

\begin{eqnarray}
<C^k>_{j^H}=& &\,{{{Z_{h}}^{-1}} \over {k!}} \oint Du(\tau)
\Biggl[{\bar g} \int_0^{\beta} d\tau u(\tau) \Biggr]^{k}
\,\nonumber \\ & & \times {}exp\Biggl[- \int_0^{\beta} d\tau {M
\over 2} \bigl( \dot{u}^2(\tau) + \omega^2 u^2(\tau) \bigr)
\Biggr]  \,\nonumber
\\
<C^k>_{j^{SSH}}& &(x)=\,{{{Z_{h}}^{-1}} \over {k!}} \oint Du(\tau)
\Biggl[-\alpha \int_0^{\beta} d\tau u(\tau) G^{\pm}[x(\tau),\tau ]
\Biggr]^{k} \, \nonumber \\ \times & & exp\Bigl[- \int_0^{\beta}
d\tau {M \over 2} \bigl( \dot{u}^2(\tau) + \omega^2 u^2(\tau)
\bigr) \Bigr]\,\nonumber
\\
\end{eqnarray}

As a direct consequence of the time retardation in the SSH {\it
e-ph} interactions, the SSH cumulant series turns out to depend on
the electron path coordinates.

In order to calculate eqs.(6), we expand the periodic oscillator
path ($u(\tau + \beta)=\, u(\tau)$) in $N_F$ Fourier components:

\begin{eqnarray}
& &u(\tau)=\,u_o + \sum_{n=1}^{N_F} 2\Bigl(\Re u_n \cos( \omega_n
\tau) - \Im u_n \sin( \omega_n \tau) \Bigr)\, \nonumber \\ &
&\omega_n=\,2\pi n/\beta
\end{eqnarray}

and choose a measure of integration which normalizes the kinetic
term in the oscillator field action:

\begin{eqnarray}
\oint Du(\tau)\equiv  & &{\sqrt{2} \over {\bigl( 2 \lambda_M
\bigr)^{2N_F+1}}} \int_{-\infty}^{\infty}{du_o}\, \nonumber \\
\times & &\prod_{n=1}^{N_F} (2\pi n)^2 \int_{-\infty}^{\infty}
d\Re u_n \int_{-\infty}^{\infty} d\Im u_n \, \nonumber \\ \oint
Du(\tau) & & exp\Bigl[-{M \over 2}\int_0^\beta d\tau
\dot{u}^2(\tau)\Bigr]\equiv 1  \, \nonumber \\
\end{eqnarray}

being $\lambda_M=\,\sqrt{\pi \hbar^2 \beta/M}$.

As $\int_0^{\beta} d\tau u(\tau)=\,\beta u_0$, using eqs.(7) and
(8), the Holstein cumulant series can be worked out analytically
as a function of the cutoff $N_F$. The first of eqs.(6) gets:

\begin{eqnarray}
& &<C^k>^{N_F}_{j^H}=\,{{{Z_h}^{-1}} \over {k!}} {{({\bar g} \beta
\lambda_M)^k (k - 1)!!} \over { \pi^{k/2} (\omega \beta)^{k+1}}}
\prod_{n=1}^{N_F} {{(2n\pi)^2} \over {(2n\pi)^2 + (\omega
\beta)^2}}\, \nonumber \\
\end{eqnarray}

where only even $k$ terms survive. As the cumulants should be
stable against the number of Fourier components in the oscillator
path expansion we set the minimum $N_F$ through the condition
$2N_F \pi \gg \omega \beta$. Since the Holstein source current
does not bear any dependence on the electron path coordinates, the
Holstein expansion in eq.(9) coincides with the cumulant series
for a time averaged electron particle path \cite{io4}.

To compute the second of eqs.(6), we expand also the electron path
under the closure condition $x(\tau + \beta)=\,x(\tau)$:

\begin{eqnarray}
& &x(\tau)=\,x_o + \sum_{m=1}^{M_F} 2\Bigl(\Re x_m \cos( \omega_m
\tau) - \Im x_m \sin( \omega_m \tau) \Bigr)\, \nonumber \\ &
&\omega_m=\,2\pi m/\beta
\end{eqnarray}

with, in general, $M_F \neq N_F$. Thus, the general cumulant term
is obtained by averaging the second of eqs.(6) over a set of
electron particle paths:

\begin{equation}
<C^k>_{j^{SSH}}=\, \sum_{x_o,\Re x_m,\Im x_m} <C^k>_{j^{SSH}}(x)
\end{equation}

Two Fourier components ($M_F=\,2$) suffice to get numerical
convergence in eq.(11). We point out that, the computation of the
anharmonic expansion in the SSH model requires for any choice of
electron path coefficients, double summations over Brillouin zone
(see eq.(4)) and inverse temperature ($\tau$) values (see the
second of eqs.(6)).

\section*{III. Phonon Thermodynamics}

The phonon thermodynamics can be derived from eq.(5) via
computation of the phonon free energy $F^{(k)}(T)=\,- \beta^{-1}
\ln \bigl[Z_{ph}[j(\tau)] \bigr]$.

To proceed one needs a criterion to find the temperature dependent
cutoff $k^*$ in the cumulant series.  We feel that, in the low $T$
limit, the third law of thermodynamics may offer the suitable
constraint to determine $k^*$. Then, set the input parameters
$\alpha$ ($g$ in the Holstein model), $J$ and $\omega$, the
program searches for the cumulant order such that phonon entropy
$E_V^{(k)}(T)$ and heat capacity $C_V^{(k)}(T)$ vanish in the zero
temperature limit. In a constant volume transformation, we compute

\begin{eqnarray}
E_V^{(k)}(T)=\,& &- \Bigl[F^{(k)}(T + 2\Delta) - 2F^{(k)}(T +
\Delta) \Bigr]/\Delta \nonumber \\ C_V^{(k)}(T)=\,& &-
\Bigl[F^{(k)}(T + 2\Delta) - 2F^{(k)}(T + \Delta) + F^{(k)}(T)
\Bigr] \nonumber \\ \times & &\Bigl({1 \over \Delta} + {T \over
\Delta^2} \Bigr)
\end{eqnarray}

$\Delta$ being the incremental step and $k^*$ is determined, at
finite $T$, as the minimum value for which the anharmonic cumulant
series converges with an accuracy of $10^{-4}$. The results are
displayed in Figures 1 for the case of a low energy oscillator and
a narrow band hopping integral, $J=\,0.1eV$. Three values of
$\alpha$ are assumed to emphasize the dependence of the SSH model
thermodynamics on the {\it e-ph} coupling. The $g=3$ value for the
Holstein model calculations yields a $\bar g$ coincident with the
largest $\alpha=\,21.74meV \AA^{-1}$.

The Holstein cumulant expansion causes a slight and slope
preserving lowering of the phonon free energy with respect to the
harmonic plot. As a consequence the Holstein entropy and heat
capacity, computed via eqs.(12), are superimposed to the harmonic
quantities. However, looking at Fig.1(d) (case $g=\,3$), we see
that a large number of cumulant terms is required to get
convergence in the Holstein series as the cutoff $k^*$ markedly
grows below $80K$, attaining the value $k^*=\,40$ at $T=\,1K$.
Thus, {\it e-ph} anharmonicities strongly renormalize the Holstein
partition function, essentially stabilize the system, but there is
no trace of anharmonicity in the phonon free energy temperature
derivatives \cite{gurevich}.

Much different is the trend of the anharmonic corrections in the
SSH model. The main features can be summarized as follows: i) the
phonon free energy (Fig.1(a)) and its derivatives (Figs.1(b), (c))
largely deviate from the harmonic plots and strongly depend on
$\alpha$; ii) there is a threshold $\alpha$ value above which the
heat capacity displays a peak. By enhancing $\alpha$ (Fig.1(c))
the height of the peak grows and the bulk of the anharmonic
effects on the heat capacity is shifted towards lower $T$; iii) at
high $T$ the anharmonic corrections renormalize downwards the
phonon free energy but their effect on the heat capacity tends to
decrease signalling that {\it e-ph} nonlinearities are rather to
be seen in the intermediate temperature range. Note that the size
of the anharmonic enhancement is about a factor $4$ larger than
the value of the harmonic oscillator heat capacity at $T \sim
\,170K$; iv) the cutoff $k^*$ in the SSH model (Fig.1(d)) attains
the value 10 at room temperature while the cumulant series
converges with a low number of terms ($k^*=\,2$) at low $T$. At
intermediate $T \in [90,270]$ , the SSH cutoff overlaps the
Holstein cutoff at the value $k^*=\,8$.

The location of the heat capacity peak along the temperature axis
strongly varies with the oscillator energy. In Fig.2 we select two
energy values and report the crossover temperatures for a set of
{\it e-ph} coupling strengths. By increasing $\omega$ at fixed
$\alpha$ the peak shifts towards high $T$ and it shows up above a
minimum $\alpha$ whose value increases at larger $\omega$. Thus,
higher energy oscillators are more stable against {\it e-ph}
induced anharmonicity since larger couplings are required to
introduce signatures of non linear behavior in the phonon heat
capacity.

So far we have focussed on the thermodynamics of the phonon
subsystem. The question arises whether and to which extent the
{\it anharmonic phonon} heat capacity affects the {\it total} heat
capacity of the SSH model. As discussed in ref.\cite{io3} the
total heat capacity is given by the phonon contribution plus a
{\it source} heat capacity which includes both the electronic
contribution (related to the electron hopping integral) and the
contribution due to the source action (the latter being $\propto
\alpha^2$, see eqs.(6) in ref.\cite{io3}). A bath of harmonic
phonons has been assumed in our previous work. Now we compare the
anharmonic phonon heat capacity ($C_V^{anh}$) and the source heat
capacity, here termed $C_V^{e-p}$ to emphasize the dependence both
on the electronic and on the {\it e-ph} coupling terms. The plot
displaying the largest peak in Fig.1(c) is reported on in Fig.3(a)
while $C_V^{e-p}$ is computed according to the method described in
in ref.\cite{io3} after setting $\alpha=\,21.74 meV \AA^{-1}$.
Also the total heat capacity ($C_V^{tot}=\,C_V^{anh} + C_V^{e-p}$)
is shown in Fig.3(a). At low temperatures, $C_V^{e-p}$ yields the
largest effect mainly due to the electronic hopping while at high
$T$, {} $C_V^{e-p}$ prevails as the source action becomes
dominant. In the intermediate  range ($T \in [90,210]$) the
anharmonic phonons provide the highest contribution although their
characteristic peak is substantially smeared in the total heat
capacity by the source term background. Fig.3(b) compares
$C_V^{anh}$ and $C_V^{e-p}$ for two increasing values of $\alpha$:
while the anharmonic peak shifts downwards (along the $T$ axis) by
enhancing $\alpha$, $C_V^{anh}$ remains larger than $C_V^{e-p}$ in
a temperature range which progressively shrinks due to the strong
dependence of the source action on the strength of the {\it e-ph}
coupling. Finally, it has to be pointed out that the low
temperature upturn displayed in the {\it total heat capacity over
{} T} ratio \cite{io3} is not affected by the inclusion of phonon
anharmonic effects which tend to become negligible at low
temperatures.

\section*{IV. Final Remarks}

The Holstein and the Su-Schrieffer-Heeger Hamiltonian present a
fundamental difference with respect to the nature of the {\it
e-ph} coupling. Mapping the real space interactions onto the time
scale we have shown that the Holstein {\it e-ph} interactions are
{\it local} in time while, in the SSH model, the electronic
hopping induces a time dependent lattice displacement thus leading
to a {\it non local} {\it e-ph} coupling. As a consequence the
source current peculiar of the SSH model depends both on time and
on the electron path coordinates. We have applied the path
integral method to analyse the perturbing effects of the source
currents (in both models) on the phonon subsystems by expanding
the phonon partition function in cumulant series. In the low
temperature limit, the constraint imposed by the third law of
thermodynamics permits to determine the cutoff $k^*$ in the
anharmonic expansion: while the SSH series converges rapidly, an
high number of cumulants is required in the Holstein model to
stabilize the low $T$ equilibrium properties. At intermediate and
high $T$, $k^*(T)$ is comparable in both models. However, looking
at the behavior of the thermodynamic properties we find striking
differences between the two models. The Holstein phonon heat
capacity does not show any trace of anharmonicity induced by the
{\it e-ph} coupling while the SSH phonon heat capacity exhibits a
peak whose location along the $T$ axis strongly depends on the
strength of the coupling $\alpha$. By increasing $\alpha$ at a
fixed $T$, $k^*$ grows and the phonon partition function becomes
larger. As a result the point of {\it most rapid decrease} in the
SSH phonon free energy versus temperature shifts downwards and the
associated heat capacity peak is found at lower $T$. At high
temperatures the free energy decreases regularly and the
anharmonic effects on the heat capacity are reduced. Infact, at
high $T$, the time ($\tau$) retardation between electron hopping
and atomic displacement is shortened as $\tau$ encounters the
upper limit of a small inverse temperature.

Finally, we have computed the general path integral in the SSH
model for an electron path moving in a bath of anharmonic
oscillators. We have derived the total heat capacity and pointed
out the contribution of the anharmonic phonon with respect to the
purely electronic and {\it e-ph} terms. The anharmonic effects are
relevant in the intermediate temperature range whereas the peak is
smeared by the electronic hopping term (on the low temperature
side) and by the source action heat capacity at high temperatures.

\begin{figure}
\vspace*{14truecm} \caption{ Phonon (a) Free Energy, (b) Entropy
and (c) Heat Capacity versus temperature in the
Su-Schrieffer-Heeger model with {\it e-ph} anharmonic
contributions due to three values of the coupling $\alpha$ in
units $meV \AA^{-1}$. The Holstein model results are computed for
the coupling $g$ (in units of $\omega$) corresponding to the
largest of the three $\alpha$ values. Also the harmonic phonon
functions are plotted for comparison. (d) Number of anharmonic
terms in the cumulant series yielding convergent thermodynamical
quantities versus temperature both in the Su-Schrieffer-Heeger
model ($\alpha$ case) and in the Holstein model ($g$ case). A low
energy oscillator is assumed. }
\end{figure}

\begin{figure}
\vspace*{8truecm} \caption{Temperature locations of the peak in
the Su-Schrieffer-Heeger phonon heat capacity versus {\it e-ph}
coupling in units $meV \AA^{-1}$. }
\end{figure}

\begin{figure}
\vspace*{8truecm} \caption{(a) Total Heat Capacity versus
temperature in the Su-Schrieffer-Heeger model. The contributions
due to anharmonic phonons ($C_V^{anh}$) and electrons {\it plus}
electron-phonon interactions ($C_V^{e-p}$) are plotted separately.
The largest $\alpha$ of Fig.1 is assumed. (b) $C_V^{anh}$ and
$C_V^{e-p}$ for two values of $\alpha$ in units $meV \AA^{-1}$.
$\omega=\,20meV$.}
\end{figure}

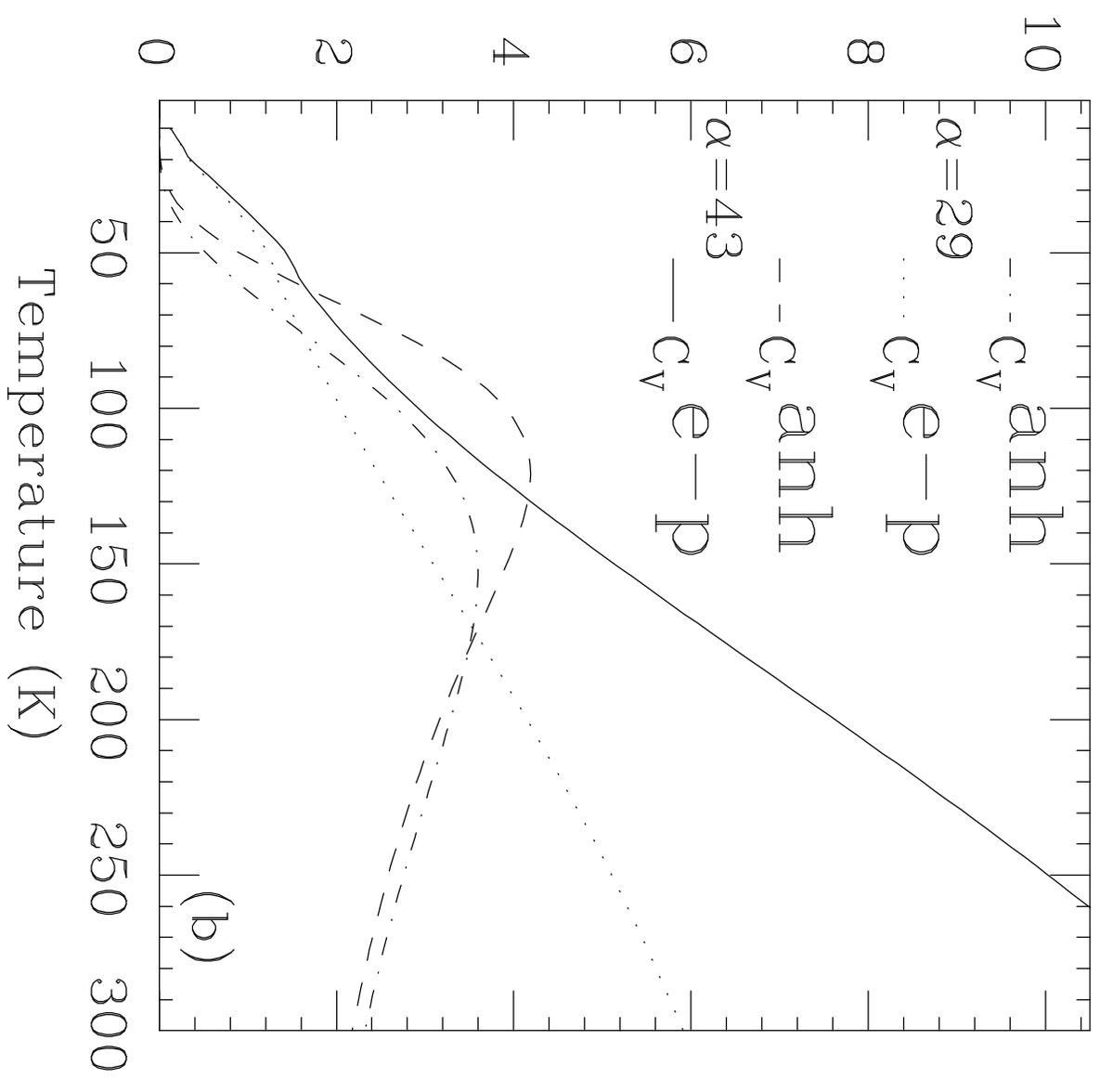
\end{document}